\documentclass[12pt]{article}
\usepackage{amsmath}
\usepackage{amssymb}
\usepackage{graphicx}
\usepackage{setspace}
\usepackage{fullpage}
\usepackage{caption}
\usepackage{subcaption}
\usepackage{indentfirst}
\usepackage{authblk}
\usepackage{multirow}

\title{Quantitative $\mu$PIV Measurements of Velocity Profiles}

\author[1,$\dagger$]{Bryant,~P.~W.}
\author[1,$\dagger$]{Neumann,~R.~F.}
\author[1,$\dagger$]{Moura,~M.~J.~B.}
\author[1]{Steiner,~M.}
\author[2]{Carvalho,~M.~S.}
\author[1]{Feger,~C.}

\affil[1]{IBM Research -- Brazil, Av. Pasteur 138 \& 146, Urca, Rio de Janeiro, CEP 22290-240, Brazil}
\affil[2]{Dept. Mech. Eng., PUC--Rio, R. Marqu\^es de S\~ao Vicente, 225, G\'avea, Rio de Janeiro, Brazil}
\affil[$\dagger$]{Equal contribution}

\date{\today}

\begin{document}
\maketitle

\begin{abstract}
In Microscopic Particle Image Velocimetry ($\mu$PIV), velocity
fields in microchannels are sampled over finite volumes within which the
velocity fields themselves may vary significantly.
In the past, this has limited measurements often to be only qualitative
in nature, blind to velocity magnitudes.
In the pursuit of quantitatively useful results, one has treated
the effects of the finite volume as errors that must
be corrected by means of ever more complicated processing techniques.
Resulting measurements have limited robustness and
require convoluted efforts to understand measurement uncertainties.
To increase the simplicity and utility of $\mu$PIV measurements,
we introduce a straightforward method, based directly on
measurement, by which one can determine the size and shape of
the volume over which moving fluids are sampled.
By comparing measurements with simulation, we verify that this method
enables quantitative measurement of velocity profiles across
entire channels, as well as an understanding of experimental uncertainties.
We show how the method permits measurement of an unknown flow rate through
a channel of known geometry.
We demonstrate the method to be robust against common 
sources of experimental uncertainty.
We also apply the theory to model the technique of Scanning $\mu$PIV,
which is often used to locate the center of a channel, and we show how and
why it can in fact misidentify the center.
The results have general implications for research and development
that requires reliable, quantitative measurement of fluid flow
on the micrometer scale and below.
\end{abstract}

\section{Introduction}
Microscopic Particle Image Velocimetry ($\mu$PIV) has become an
invaluable measurement
technique for probing fluid flow in systems at the micro- and the nano-scale~\cite{williams2010advances,lindken2009micro,wereley2010recent,adrian2010particle}.
Well-controlled and understood $\mu$PIV experiments performed on
increasingly complex systems will become crucial to maintaining the rapid
progress observed in the biomedical and natural resources
industries~\cite{sackmann2014present,mark2010microfluidic,gunda2011reservoir}.
When imaging only over a region small compared to the size of a
microfluidic system, it is sometimes possible optically to restrict the volume
over which the system is probed.
The result is often a quantitative measurement of fluid flow in that
region~\cite{joseph2005direct,tsai2009quantifying}.
To image simultaneously as many flow features as possible, however,
one requires a field of view of size comparable
to the length scales of the system.
Unfortunately, this leads inevitably to having also a
large depth of field along the optical axis,
which is an issue~\cite{meinhart1999piv,
sott2013mupiv,kloosterman2011flow} that hinders a straightforward quantification
and interpretation of results. 
For $\mu$PIV to reach its full potential as an investigative and
design tool, it must provide repeatable and quantitatively useful results.
It has been an open question whether or not this is possible
for $\mu$PIV with a depth of field of size comparable to that of the
microfluidics system it measures.

There are two standard approaches to treating the effects of a large
depth of field.
The first is to limit $\mu$PIV to qualitative comparisons of velocity profile 
shapes, often away from channel walls.
In such cases, one treats the volumetric flow rate of the fluid as a free parameter
used to match expected profiles to the measured data~\cite{meinhart1999piv,
sott2013mupiv}.
In the second approach, one treats finite volume effects as measurement errors,
particularly near the walls, and attempts to remove as many of these effects as
possible via data processing.
This approach has, to some extent, enabled the quantitative evaluation of $\mu$PIV
data~\cite{kloosterman2011flow,rossi2012effect,westerweel2004single,
cierpka2012particle}.

A shortcoming of the latter method, however, is the often complicated
and convoluted data processing that limits or even precludes the physical
understanding of the fluid flow and its analysis.
It bears the risk of over-fitting results to the extent that important flow
features are missed, or even
misdiagnosing the causes of observed phenomena.
An example is the spurious, increasing velocity near the channel walls that 
sometimes appears~\cite{rossi2012effect,cierpka2012particle}.
In most cases noisy or uncertain experimental data are excluded from
the measured flow profile altogether, without physical justification.
A related problem is that excessive data processing introduces additional
sources of experimental uncertainty in an obscure fashion.

In this article, we investigate the results of a series of experiments performed
with a large depth of field $\mu$PIV, in which we image across the entire
microchannel and measure the velocity profile from wall to wall.
Rather than use data processing to remove the effects of finite
measurement volumes, we introduce the concept of Sampling Volume (SV)
over which the velocity field can be averaged to recover the measured
velocity profiles.
We then systematically determine the size and shape of the SV using an
error minimization procedure.
For our experiment, which is described below, we find that measurements near
the walls reflect in a straightforward manner the non-trivial intersection of
the SV with the edges of the microchannel.
Consequently, the entire measured profile contains information useful for
determining the SV, and one should therefore not discard features measured near
channel walls.

We begin with a discussion of $\mu$PIV for large depth of field
and a motivation for our definition of the SV.
We then describe our experimental setup and how our data
are processed and analyzed.
We also discuss the expected velocity field in the channel,
and its relevant sources of uncertainty.
Then we show how averaging the velocity field
over a rectangular SV reproduces to within our experimental uncertainty the
measured velocity profiles across the entire channel.
We also demonstrate how, because of the simplicity of our method and the
minimal post-processing of data, $\mu$PIV can in some cases be used to 
infer an unknown velocity field or flow rate in a channel of known
cross section.
We close with a series of use cases for our approach, which include
robustly measuring velocity profiles in the presence of camera misalignment
and very noisy data that would typically be deemed useless,
a demonstration of how data processing affects
the SV, and finally a critical analysis of the procedure known as
``Scanning PIV,'' which is often used to place a measurement domain at the
center of microchannels~\cite{kloosterman2011flow,chayer2012velocity}.
We show why, for finite-sized SV, Scanning PIV does not necessarily
locate the center of a microchannel, and should therefore be used with caution.

\section{Particle Image Velocimetry}

\subsection{General description}
Particle Image Velocimetry (PIV) is an optical method for visualizing and
measuring flow fields.
One seeds a fluid with tracer particles, and
if the particles follow the fluid flow without disturbing
it significantly, by analyzing their trajectories one can infer the
velocity field of the moving fluid.
To determine the particle displacements one uses consecutive snapshots
of the system taken at times $t$ and $t+\Delta t$, containing several particle
images at various positions.
The snapshots are subdivided into smaller regions called ``interrogation windows''
and the net particle displacement is calculated for each one of them.
Finally, the velocity representative of each window is calculated simply by
dividing the respective displacement by the interval, $\Delta t$, between frames.

Analyzing and interpreting a PIV measurement is complicated for the following
reasons.
First, the velocity obtained from this process is an average in space and time,
and its accuracy depends on the flow field itself,
as well as on both spatial and temporal resolution of the experimental
setup~\cite{westerweel1997fundamentals}.
Second, tracer particles are susceptible to Brownian motion, and as a
consequence their positions in images must be correlated within sub-ensembles
to suppress the effect of random movements and determine net particle
displacements.
The application of cross-correlation techniques ultimately provides a 2D
projection of the 3D velocity field of the fluid located within the imaged
region, as viewed from the camera~\cite{adrian2010particle}.

\begin{figure}[htp]
\centering
\includegraphics[width=\textwidth]{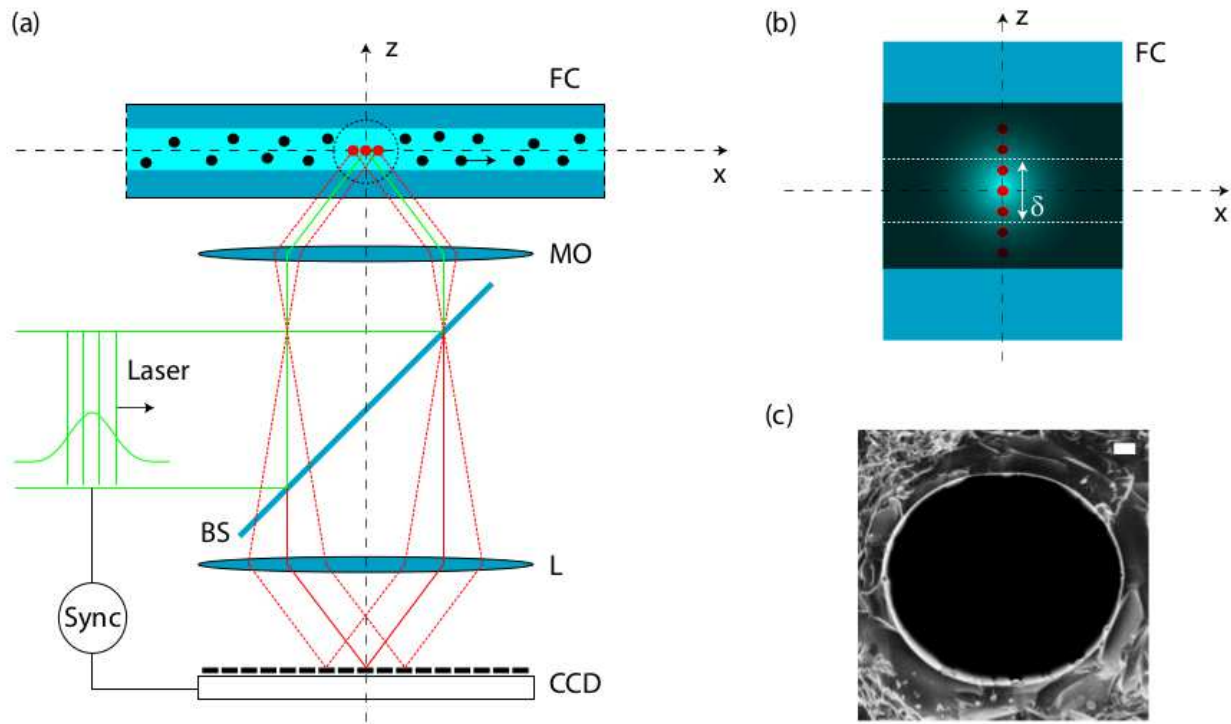}
\caption{(a) Simplified schematic of the $\mu$PIV setup.
The flow direction of fluorescent particles in the fluidic channel (FC) is
indicated by an arrow.
MO: microscope objective; BS: dichroic beam splitter; L: lens; CCD: charge
coupled device array detector.
(b) Schematic of the fluidic channel showing the depth of the SV, $\delta$.
(c) Scanning electron microscope image of the cross section of the fluid channel
used in the flow experiments.
The length of the white scale bar represents 10 $\mu$m.}
\label{fig:pivsup}
\end{figure}

In Microscopic PIV ($\mu$PIV), see Figure~\ref{fig:pivsup}(a),
fluorescent tracer particles in a fluid channel of arbitrary geometry
are observed by means of an optical microscope setup. 
The size of the tracer particles needs to be chosen carefully such that 
particles emit a sufficient amount of light for detection without 
perturbing the flow.
To perform the measurement, one focuses a laser pulse on the fluid
channel to excite the tracer particles. 
The particles then emit light at a red-shifted fluorescence wavelength that is
collected by the same objective lens, alongside with the reflected laser light. 
The reflected excitation laser light is then filtered out by a dichroic beam 
splitter while the fluoresced light passes, rendering fluorescence images of
spatially resolved particles on the CCD array. 
As indicated in Figure~\ref{fig:pivsup}(a), fluorescent particles (red)
in the object plane are spatially resolved in the conjugate image plane with a 
lateral resolution in $x$ and $y$ that is determined by the specifics
of the optical setup.
However, a particle's image will only be detectable if its $z$-position
falls within the finite range labeled $\delta$ in Figure~\ref{fig:pivsup}(b),
which depends not only on the characteristics of the optical setup but also on
the data processing algorithms used in the analysis.
In the following section, we will introduce a concept for the determination
of the volume probed by a $\mu$PIV experiment.

\subsection{Sampling Volume}\label{sec:intrsv}
In this paper,
our main concern is a well-known complication inherent to $\mu$PIV
with a large depth of field, for which tracer particles are imaged over a finite
volume with a characteristic length scale
on the order of the scale of the microchannels themselves.
Besides the well established \textit{depth of field} and \textit{depth of
focus}~\cite{adrian2010particle,novotny2012principles}, several other quantities
have been used to quantify the depth over which the system is probed.
Among them are
the laser light sheet~\cite{adrian2010particle,meinhart1999piv}, the depth of
measurement~\cite{meinhart2000volume}, and the depth of 
correlation~\cite{sott2013mupiv,kloosterman2011flow,olsen2000out}.
Although they have various meanings and account for various effects,
many of them derive from the basic definition of the depth of field.

The term ``laser light sheet''~\cite{adrian2010particle,meinhart1999piv} was
introduced to denote the illuminated region inside the fluid system.
In most $\mu$PIV applications the light sheet and the entire fluid system are of
comparable sizes. 
When that is the case, all particles inside the channel are excited by the
laser light and, in general, all fluorescent particles within the $(x,y)$ field
of view of the optical setup and within the depth of field, $\Delta z$,
create in-focus images on the CCD detector.
The geometrical shape of this volume can be approximated as a parallelepiped,
and the focal $(x,y)$-plane oriented perpendicular to the optical $z$ axis
cuts this volume in half. 

In standard far-field fluorescence microscopy the total point spread function
and, hence, the size of the focal volume is determined by the product of the
excitation point spread function and the detection point spread
function~\cite{novotny2012principles}.
In $\mu$PIV, however, where image processing is used to determine the particles'
displacements over successive frames, there are additional contributions that
alter the effective volume that is being probed, such as the application of
the image pair correlation function~\cite{meinhart1993parallel}. 
As a consequence, a ``depth of correlation'' (DOC) was introduced to 
estimate the depth of the volume that contains those particles contributing
significantly to the spatial correlation
function~\cite{sott2013mupiv,kloosterman2011flow,olsen2000out}.
The calculated DOC is useful for identifying the experimental parameters
relevant to the probed volume.
However, it is based on an assumed intensity threshold
beyond which imaged particles do not ``cause a
significant bias in the velocity estimate''~\cite{cierpka2012particle}.
It is thus not quantitatively well defined, and has not led to conclusive
results in comparison with experiments~\cite{sott2013mupiv,kloosterman2011flow}.
In Section~\ref{sec:fitcompar} we will present a series of results
that indicate the DOC should be used only as an order-of-magnitude estimate
of the probed volume.

Despite significant effort to determine precisely the probed volume,
a simultaneously intuitive and quantitative definition is lacking.
To address this issue, we introduce a quantity with a
straightforward and verifiable interpretation.
We call it Sampling Volume (SV), and it is designed to account for the
combined effects of illumination scheme, optical depth, and image processing.
The concept, as opposed to previously defined quantities such as the DOC, is
strikingly simple: the SV is defined as the volume over which one 
averages the velocity field of the fluid in order to recover
the measured profile.

We begin with a complete description of our averaging process for one experiment
performed on a simple, straight channel, for which we will ignore any variation
in measured velocity down the channel ($x$ direction).
We denote the velocity field within the channel by
$\vec{v}(x,y,z)$.
Because of the translational symmetry in $x$, we take
$\vec{v}(x,y,z)\rightarrow \vec{v}(y,z)$.
The velocity averaged over the SV, which is compared with measurements, is then
\begin{equation} \label{eq:vavedoc}
\langle\vec{v}\rangle_{SV}(y)=\frac
{\int_{z_{\mbox{\tiny --}}(y)}^{z_{\mbox{\tiny +}}(y)}\,dz\,\vec{v}(y,z)}
{z_{\mbox{\tiny +}}(y)-z_{\mbox{\tiny --}}(y)},
\end{equation}
as a function of $y$, which is the direction across the channel, as
seen below in Figure~\ref{fig:cxFit}.
In~\eqref{eq:vavedoc}, the integration bounds for a rectangular SV
are defined by
\begin{equation} \label{eq:bounda}
z_{\mbox{\tiny --}} \equiv \textnormal{Max}
\left(c-\frac{\delta}{2},z_{\textnormal{bot}}(y)\right)
\end{equation}
and
\begin{equation} \label{eq:boundb}
z_{\mbox{\tiny +}} \equiv \textnormal{Min}
\left(c+\frac{\delta}{2},z_{\textnormal{top}}(y)\right).
\end{equation}

In~\eqref{eq:bounda} and~\eqref{eq:boundb}, $\textnormal{Max}(\cdot)$ and
$\textnormal{Min}(\cdot)$ denote the maximum and minimum values
of the arguments.
The bottom and top locations of the channel itself are $z_{\textnormal{bot}}(y)$
and $z_{\textnormal{top}}(y)$, respectively.
The vertical center of the SV, as measured with respect to the channel center,
is denoted by $c$, and the vertical depth, or thickness, of the SV
is represented by $\delta$.
The selection of the maximum or minimum values in the integration bounds
is necessary to keep from continuing the average beyond the channel's
extent, where no fluid exists.
In Section~\ref{sec:prevwork} is a demonstration of how failing to
cut off the average properly at the edge of the channel can lead to a
calculated velocity of limited applicability.
Throughout Sections~\ref{sec:results} and~\ref{sec:applications} are
several demonstrations of the success of using
Equation~\eqref{eq:vavedoc} to explain measurements.

\section{Experimental Methods and Data Analysis} \label{sec:expsetup}

\subsection{Flow Field Measurements}
In this section, we briefly describe the experimental setup used for 
performing all the fluid flow experiments discussed in this article. 
Further description of the experimental setup can be found 
in Ref.~\cite{gutierrez2013flow}.
The experiments were performed using a commercially available 
$\mu$PIV system (TSI Inc.), composed of an optical microscope, CCD camera, 
two lasers, and a laser pulse synchronizer. 
The system is operated by a data acquisition and management software 
(Insight 3G, TSI Inc.). 
The inverted microscope (IX71S1F-3, Olympus) is equipped with a 10x/0.30 
air objective (UPlanFL~N, Olympus).
The Peltier-cooled CCD (Sensicam 630166, PowerView) has a resolution of 
1376 $\times$ 1024 pixels (1.4MP) and a 2x projection lens. 
The two pulsed Nd:YAG lasers (Gemini PIV-15, NEW WAVE) operate at 532 nm 
and a frequency of 15 Hz, providing a pulse energy of 50 mJ.
From the specifications of the optical setup we can calculate our depth of
field~\cite{novotny2012principles} to be approximately 20\,$\mu$m.
The laser pulse synchronizer (610034, TSI Inc.) controls the timing sequence 
of laser pulses and exposure time of the CCD. 
For each flow experiment, 100 pairs of images were collected with a time 
interval, $\Delta t$, of 500\,$\mu$s between successive images. 

A glass microfluidic device manufactured by Dolomite Centre Ltd, 
UK was used to perform the flow experiments. 
The device has a straight channel with an elliptical cross-section, 
with a semi-minor axis of 50\,$\mu$m and semi-major axis of 55\,$\mu$m. 
To visualize the flow we use a mix of 14\% aqueous solution of fluorescent particles of 1\,$\mu$m in diameter 
(ex/em 542/612 nm, Thermo Scientific Fluoro-Max Dyed Red Aqueous Fluorescent Particles) and 
86\% purified water (Thermo Scientific Nalgene Analytical Filter, 0.2\,$\mu$m).
This fluid mix is injected in the device with 
the use of a syringe pump (11 Elite syringe pump 704501, Harvard Apparatus). 
The injection rates ranged from 75\,$\mu$l/h to 100\,$\mu$l/h.

\subsection{Measurement Uncertainty}\label{sec:uncquant}
The extent to which we can determine the size and shape of our SV depends
on our experimental uncertainties, as will be demonstrated in
Sections~\ref{sec:results} and~\ref{sec:applications}.
We therefore performed several calibration procedures to assess the accuracy
and precision of our measurements.

The major sources of uncertainty in the fluid flow itself are in the
microchannel geometry and the flow rate provided by the syringe pump. 
With a scanning electron microscope, we imaged representative cross sections of
our channels.
Figure~\ref{fig:pivsup} contains a sample image.
We digitized the boundary from two separate images and found that all
digitized points lie within the region between two ellipses with
semi-major and semi-minor axes
of $(55\pm 1.5)\,\mu\textnormal{m}$ and $(50\pm 1.5)\,\mu\textnormal{m}$,
respectively.
The syringe pump was calibrated by measuring the amount of fluid injected
over time. 
Five measurements were taken for each of three injection rates. 
The measurements were also compared with the total infused volume and the
time elapsed, as given on the pump's display.
We obtained a 3\% uncertainty in the volumetric flow rate,
which is consistent with the uncertainty of 0.5\% provided by the manufacturer.
For our results in Sections~\ref{sec:results} and~\ref{sec:applications},
we assume an uncertainty of 0.5\% of the flow rate.
We also tested the syringe pump connected to the microfluidic system
to check for internal pressure effects. 
Again we measured the infused volume over time.
No significant change in the injection rates was observed.

To address the uncertainty in lengths measured by the $\mu$PIV system,
we calibrated the camera's field of view using a standard calibration target.
For our experimental setup the field of view is
(440$\pm$5)\,$\mu$m $\times$ (325$\pm$5)\,$\mu$m,
in agreement with the equipment's specifications.

\subsection{Data Processing}\label{sec:datproc}
\begin{figure}[htp]
\centering
\begin{subfigure}[t]{\textwidth}
    \caption{\hfill~}
    \includegraphics[width=\textwidth]{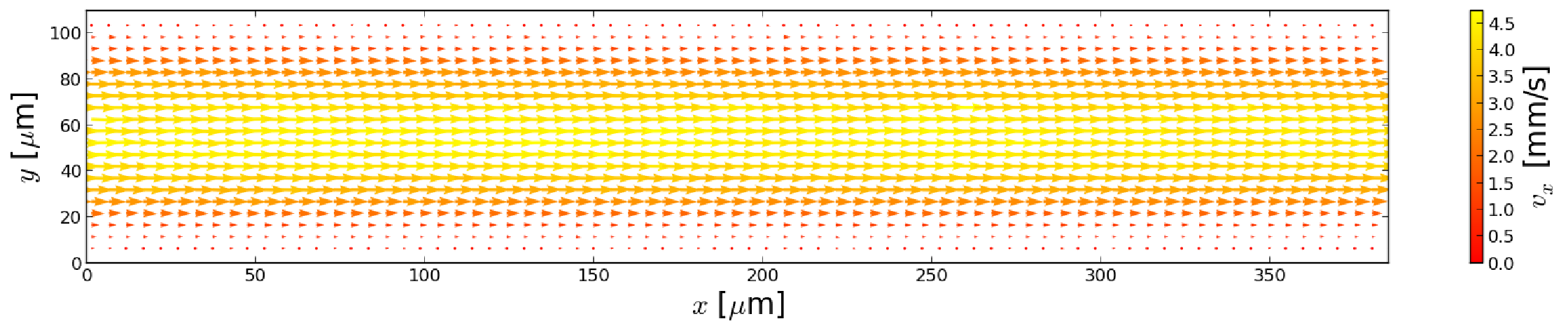}
    \label{fig:quiver_I3G}
\end{subfigure}
\begin{subfigure}[t]{\textwidth}
    \caption{\hfill~}
    \includegraphics[width=\textwidth]{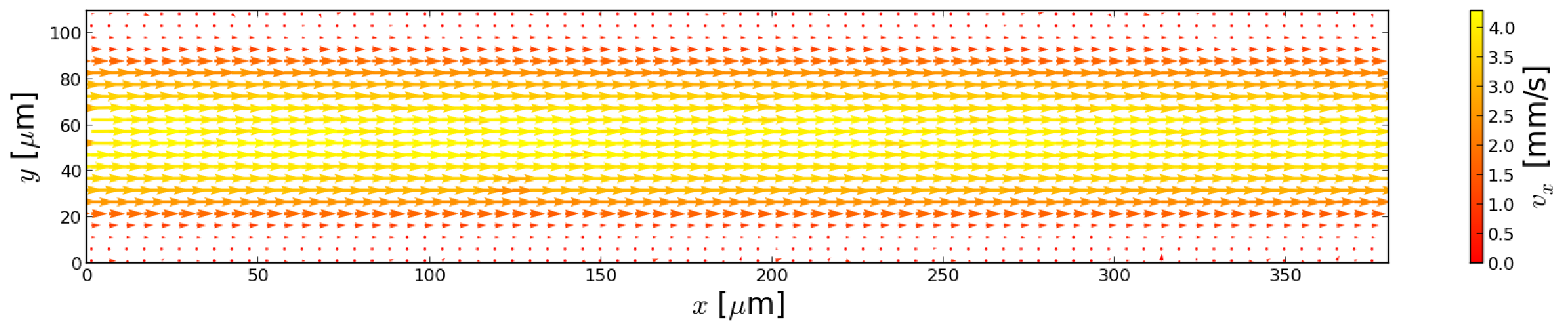}
    \label{fig:quiver_OPIV}
\end{subfigure}
\caption{Quiver plots of the measured velocity field as processed by
(\subref{fig:quiver_I3G}) Insight 3G and (\subref{fig:quiver_OPIV}) OpenPIV.}
\label{fig:quiver}
\end{figure}

Image processing of the results presented in Section~\ref{sec:results} was
carried out using the Insight 3G software, starting from the raw images.
We used an interrogation window of 32 $\times$ 32 pixels, with a 50\% overlap
according to the Nyquist sampling criterion.
The FFT correlator and the 3-point Gaussian peak detection were employed in the
determination of particle displacements.
In order to increase the signal-to-noise ratio, we subtracted background noise
and performed an ensemble average over correlation functions across 100 pairs of
snapshots. 
From the camera's pixel resolution and the field of view we know that one
pixel corresponds to 0.32\,$\mu$m in both the $x$ and the $y$ direction.
The overall resolution is therefore 5.12\,$\mu$m.
With this resolution we obtain 85 velocity vectors along the channel length and
21 along the channel width.
Because of the low Reynolds number, 
the resulting flow is as expected laminar and stable.
Figure~\ref{fig:quiver}(a) shows a typical velocity profile, which will
be further analyzed in Sec.~\ref{sec:results}.

Velocity profiles were also obtained independently
starting from the same raw images and analyzing them with the
OpenPIV~\cite{taylor2010long}
software~\footnote{\tt http://www.openpiv.net}.
The processing options for OpenPIV, such as the geometry of the
interrogation windows, the cross-correlation method, and the peak
detection algorithm, were the same as those employed by Insight 3G.
The difference was in the noise reduction algorithms: instead of
performing an ensemble average over correlation functions, we averaged the flow
field themselves over the 100 pairs of snapshots, and we did not perform 
background noise subtraction.
The results of the two processing methods are discussed in
Section~\ref{sec:fitcompar}.

\section{Results} \label{sec:results}
\begin{figure*}[htp]
\centering
\begin{subfigure}[t]{0.48\textwidth}
    \caption{\hfill~}
    \includegraphics[width=\textwidth]{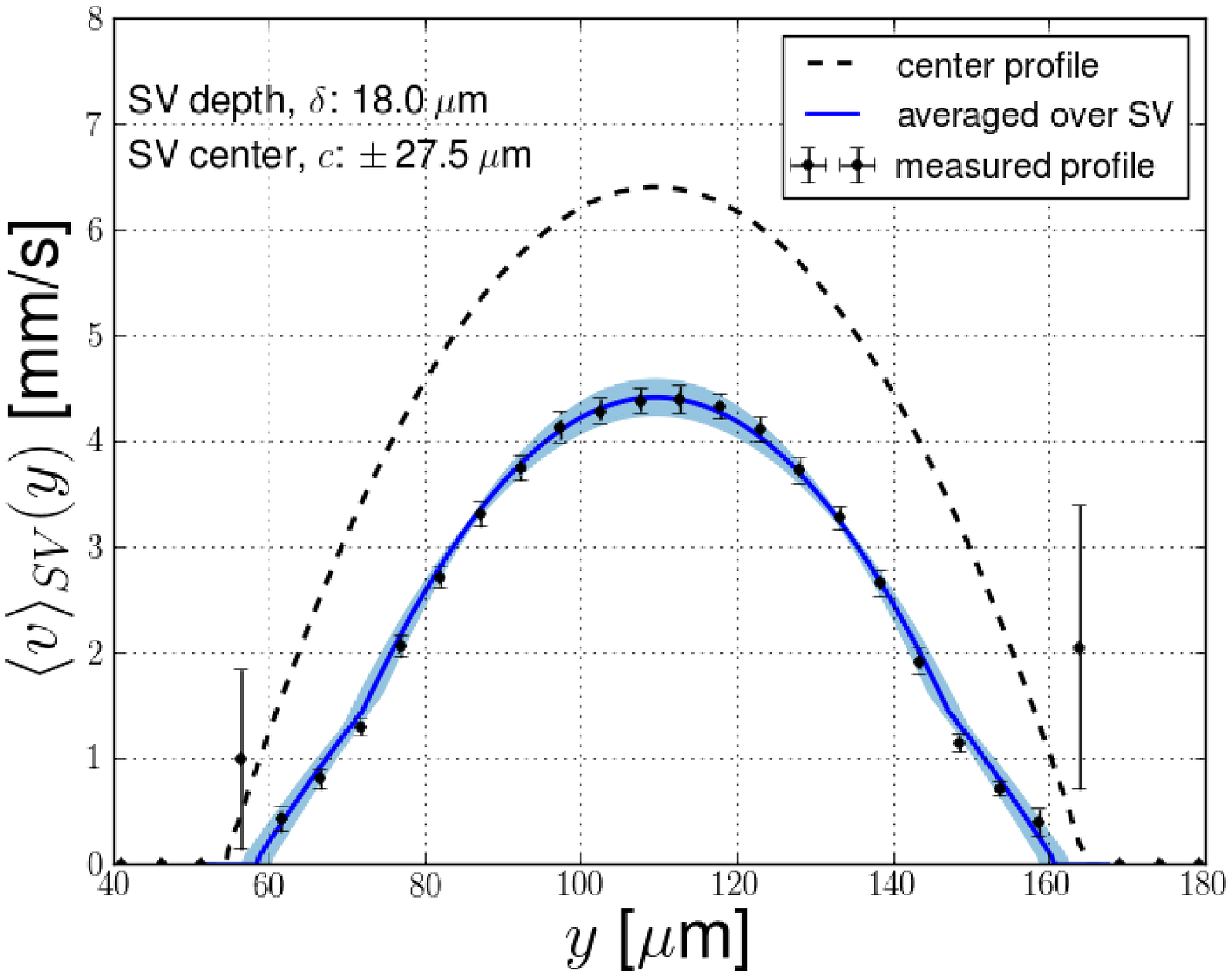}
    \label{fig:cxfit1}
\end{subfigure}
\hfill
\begin{subfigure}[t]{0.48\textwidth}
    \caption{\hfill~}
    \includegraphics[width=\textwidth]{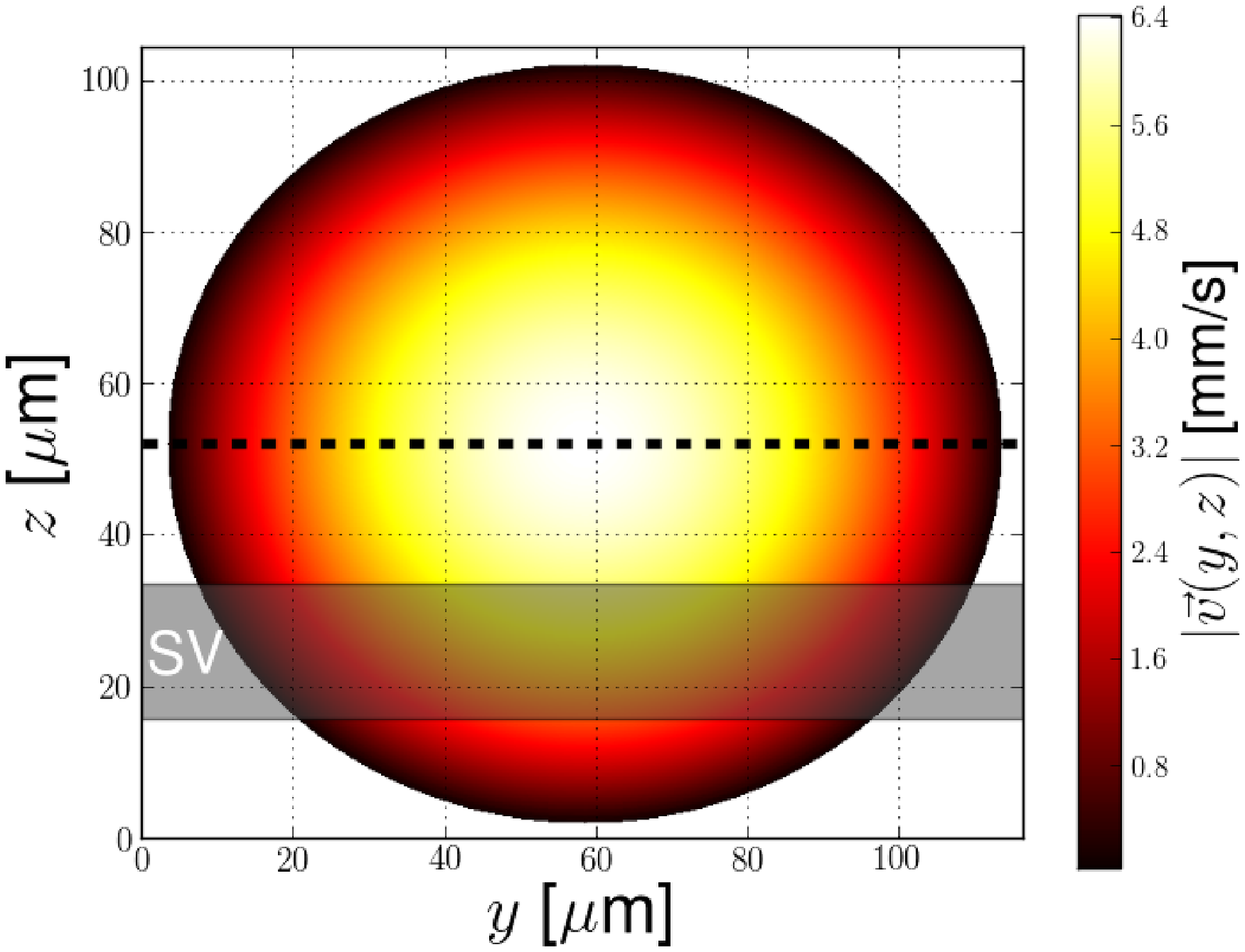}
    \label{fig:cxfit2}
\end{subfigure}
\caption{(\subref{fig:cxfit1}) Measured and theoretically expected velocity
profiles for an injection rate of 100\,$\mu$l/h.
(\subref{fig:cxfit2}) The magnitude of the simulated velocity field,
$|\vec{v}(y,z)|$, is overlaid by the SV (shaded rectangle)
that results in the fit.
The dashed line in both images corresponds to a centered SV of zero thickness.}
\label{fig:cxFit}
\end{figure*}

For comparison with theory we averaged the profiles located along
the (streamwise) $x$ direction.
The resulting average profile for an injection rate of 100\,$\mu$l/h
is pictured in Figure~\ref{fig:cxFit}(\subref{fig:cxfit1})
as dots with vertical error bars corresponding to one standard deviation.
Horizontal error bars representing the uncertainty in length calibration
discussed in Section~\ref{sec:uncquant} are too small to be seen.
The large speeds with large error bars beyond the channel walls are numerical
artifacts from data processing.

We used Equation~\eqref{eq:vavedoc} to calculate $\langle v\rangle_{SV}(y)$
for a given thickness, $\delta$, and position, $c$, of the SV.
To compute the velocity field, $\vec{v}(x,y,z)$, we used the
Lattice Boltzmann Method~\cite{chen1998lattice}, to which is input the
injection rate of the pump, the channel geometry, and the fluid properties.
To determine the position, $c$, and the thickness, $\delta$, of the SV
that results in the measured profile, we searched for possible fits to
the data by varying $c$ and $\delta$ while keeping $\vec{v}(x,y,z)$ fixed.
In other words, we did not vary the pressure drop or the injection flow
rate, and thus the magnitude of the theoretically expected velocity field
is not a free parameter.
In Figure~\ref{fig:cxFit}(\subref{fig:cxfit1}),
the solid blue line is $\langle v\rangle_{SV}(y)$ calculated
for an elliptical channel with semi-major and semi-minor
axes of $55\,\mu\textnormal{m}$ and $50\,\mu\textnormal{m}$, respectively,
and a flow rate delivered by the pump of 100\,$\mu$l/h.
The shaded area represents the combined effect
of the uncertainty in the channel shape and the flow rate,
as discussed in Section~\ref{sec:uncquant}.
The dashed line, included for reference, is the calculated velocity profile
at the center of the channel, $c=0$.

\begin{figure}[htp]
\begin{center}
\includegraphics[width=.5\textwidth]{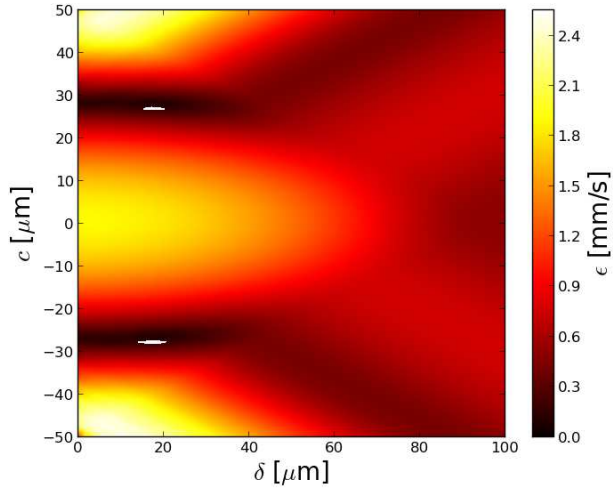}
\end{center}
\caption{Dependence of the residual, $\epsilon$, with respect
to the fit parameters $c$ (the SV center) and $\delta$ (the SV thickness).
The white spots around the minima depict regions in which $\epsilon$ is no
more than 20\% larger than its lowest value.}
\label{fig:fitpar}
\end{figure}

The fit in Figure~\ref{fig:cxFit} is for
$\delta=18\,\mu\textnormal{m}$ and $c=\pm 27.5\,\mu\textnormal{m}$.
The position can be positive or negative because the channel has a reflection
symmetry with respect to a horizontal plane bisecting it.
The lower location is depicted in Figure~\ref{fig:cxFit}(\subref{fig:cxfit2}).
These two parameters that determine the disposition of the SV
were found by sweeping through a range
of values, for which Figure~\ref{fig:fitpar} shows the residual, $\epsilon$.
It is clear that no other choices for the center and depth of a rectangular SV
can fit the data as well.

Based on this fit, we conclude that, to within our experimental
uncertainty, for this test there is a rectangular SV over which one can
average the velocity field to match the measured velocity profile.
The result is a quantitatively useful $\mu$PIV measurement of a velocity
profile across the entire width of a channel.
In Section~\ref{sec:applications} are several more examples of successful fits
to experimental results.
It has been claimed recently~\cite{kloosterman2011flow}, however,
that the simple average over a sampling region cannot reproduce velocity
profiles obtained by $\mu$PIV with a large depth of field.
Our analysis in Section~\ref{sec:prevwork} indicates that this claim
may result from the incorrect use of a technique called Scanning PIV to
locate the vertical center of a channel.

\section{Inferring Flow Rate, Channel Geometry, or Velocity Field}
In Section~\ref{sec:results}, we demonstrated how to determine the
SV for a system in which the flow rate, channel geometry, and fluid properties
are known.
An important benefit of minimizing the post-processing of data and instead
working with the SV is that one may be able to reverse the analysis and
infer either the flow rate, the velocity field, or
the channel geometry in systems where one of them is not known.
For example, a common approach when one uses $\mu$PIV to measure the
flow rate of, for example, blood in an artery or oil in porous
media~\cite{gunda2011reservoir}, is to measure the velocity profile
and then match the maximum
velocity to the peak of the theoretically expected profile,
such as the Hagen-Poiseuille profile for cylindrical channels.
As is clear from our analysis of the SV, however,
this approach will underestimate the experimental flow rate.

One can already obtain a better evaluation from the same data by acknowledging
the presence of the SV.
If one first calibrates the $\mu$PIV apparatus for a well-controlled
experiment in which the thickness, $\delta$, of the SV is determined for a 
given set of optical and processing parameters, as described in
Section~\ref{sec:results}, one can use the
knowledge of $\delta$ to understand the degree to which a measured velocity
profile will underestimate the actual flow rate.

\begin{figure}[htp]
\centering
\includegraphics[width=.5\textwidth]{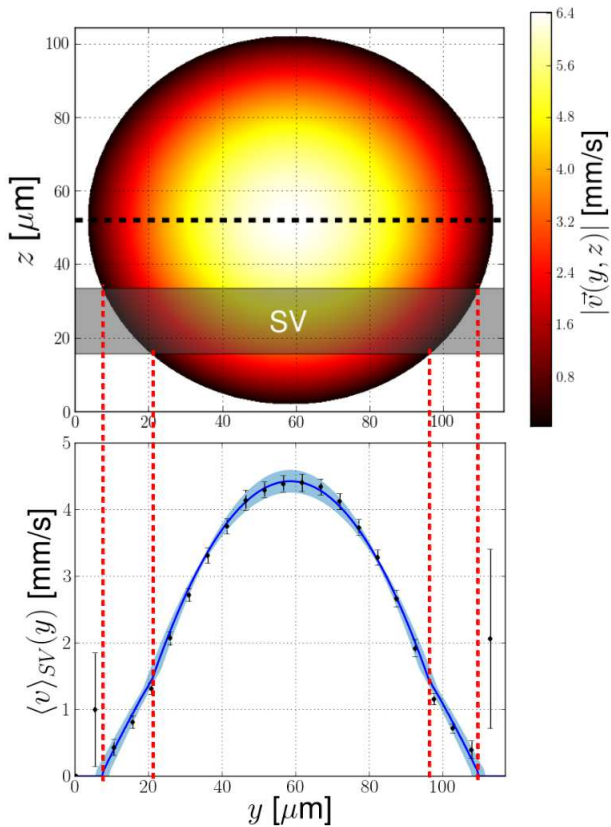}
\caption{Correspondence between the intersections of the channel walls 
with the SV and the position of the near-wall features.}
\label{fig:revProc}
\end{figure}

Perhaps more interestingly, by retaining all measured features, such as the
kinks near the walls in the profiles seen in
Figure~\ref{fig:cxFit}(\subref{fig:cxfit1}),
for some channel shapes one can determine with a single experiment
the size and position of the SV, without fitting $\langle v\rangle_{SV}(y)$
to the measured profile.
As shown in Figure~\ref{fig:revProc} for our experiment,
the $y$ values where the measured profile goes to zero uniquely determine where
the top of the SV intersects the channel's walls, where the velocity
field itself is zero.
Similarly, the profile's kinks
determine where the bottom of the SV intersects the walls.
With $c$ and $\delta$ uniquely determined, and given the measured velocity
magnitudes, one can infer the flow rate
or even the velocity field of the flowing fluid.
For example, if we had known the channel geometry but not the flow rate
for the measurement described above,
we could have recovered, to within experimental
uncertainty, the entire velocity field by performing several simulations
with various flow rates.

In this scenario, one must know the fluid properties and have enough
knowledge of possible flow rates to choose correctly the $\mu$PIV parameters
such as the $\Delta t$ between images.
For some channel shapes and SV sizes, a unique determination of $c$
and $\delta$ may not be possible from a single measured profile.
In that case, the theory can still be used to identify a set of possible
solutions and to judge the relative likelihood of the possibilities.
Furthermore, by parallel reasoning, one can reverse the process:
If one knows the flow rate but not the channel geometry,
one can in principle iterate over possible channel shapes to infer
as much as possible about the unknown geometry of the microchannel itself.

Another intriguing possibility is to verify flow models for
complex fluids that present non-Newtonian behavior, for which theoretical
models or numerical solutions may not be accurate.
In this case, one must either determine the SV as in Figure~\ref{fig:revProc},
or else by means of a calibration experiment using known optical parameters
and a well-understood fluid, such as we have described
in Section~\ref{sec:results}.
Then one can measure the velocity profile of the complex fluid, and the
result will be a $\mu$PIV measurement that can provide quantitative
insight into the various theoretical treatments of the complex fluid.

\section{Use Cases}  \label{sec:applications}
Here we describe several use cases for the
simple theory of the SV presented in Section~\ref{sec:results}.
These use cases demonstrate its robustness against
common sources of experimental error and against noisy data,
its utility for quantifying the effect of processing variables, and
how the theory has led to a deeper understanding of the Scanning PIV technique.

\subsection{Robustness Against Camera Misalignment}
\begin{figure}[htp]
\begin{center}
\includegraphics[width=.5\textwidth]{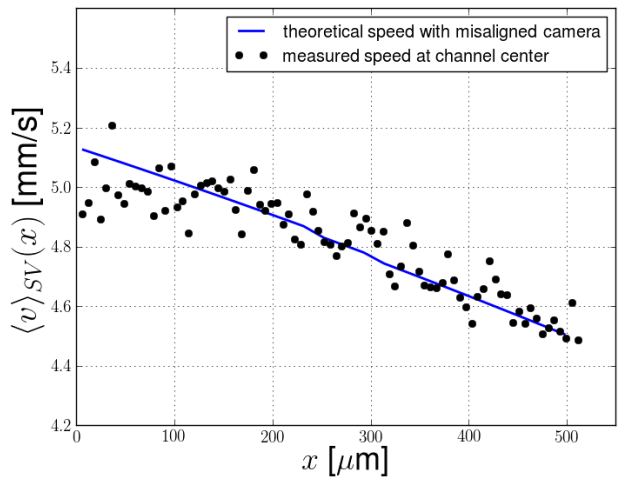}
\end{center}
\caption{The points represent the speed measured at the center of the cross
section for several positions along $x$, the direction of flow.
The solid line shows the theoretically expected speed,
$\langle v\rangle_{SV}(x,y=y_{\textnormal{center}})\equiv \langle v\rangle_{SV}(x)$
calculated at the center of the profile using
Equations~\eqref{eq:cxX},~\eqref{eq:boundax}, and~\eqref{eq:boundbx}.}
\label{fig:fitDC}
\end{figure}

In some experiments, we measured a consistent and smooth drop in
the velocity along the $x$ direction, which is the direction of flow.
One example can be seen in Figure~\ref{fig:fitDC}, for the speed at
the center (in $y$) of the channel as it changes in the $x$ direction.
Once an experiment was prepared and fluid was flowing, the change in velocity
down the channel was consistent, even for other
locations along the length of the channel.
In other experiments, such as the one presented in Figure~\ref{fig:cxFit},
there was no appreciable change in velocity down the channel.
The simplest explanation for this occasional smooth change in velocity is an
occasional misalignment between the camera and the microchannel.

With our method, one can easily treat a misaligned camera.
The translational symmetry of the channel is unaffected,
so we continue to use $\vec{v}(x,y,z)\rightarrow \vec{v}(y,z)$
for the velocity field inside the channel.
The position and shape of the SV do depend on the
optical system, however, so we model a misalignment of
the camera by varying in $x$ the position of the center of the SV:
$c\rightarrow c(x)$.
The result is an expected measured velocity that also varies in $x$,
$\langle\vec{v}\rangle_{SV}(x,y)$, which is calculated as
\begin{equation} \label{eq:cxX}
\langle\vec{v}\rangle_{SV}(x,y) = \frac
{\int_{z_{\mbox{\tiny --}}(x,y)}^{z_{\mbox{\tiny +}}(x,y)}\,dz\,\vec{v}(y,z)}
{z_{\mbox{\tiny +}}(x,y)-z_{\mbox{\tiny --}}(x,y)}.
\end{equation}
Here the integration bounds must be defined as
\begin{equation} \label{eq:boundax}
z_{\mbox{\tiny --}}(x,y) \equiv \textnormal{Max}
\left(c(x)-\frac{\delta}{2},z_{\textnormal{bot}}(y)\right)
\end{equation}
and
\begin{equation} \label{eq:boundbx}
z_{\mbox{\tiny +}}(x,y) \equiv \textnormal{Min}
\left(c(x)+\frac{\delta}{2},z_{\textnormal{top}}(y)\right).
\end{equation}

For the fit in Figure~\ref{fig:fitDC} the thickness of the SV, $\delta$,
is a constant $18\,\mu\textnormal{m}$.
The center of the SV, $c(x)$, varies linearly from
$c(x=6\,\mu\textnormal{m})=-21.7\,\mu\textnormal{m}$ to
$c(x=517\,\mu\textnormal{m})=-27\,\mu\textnormal{m}$.
The channel depth is $100\,\mu\textnormal{m}$, so the fit in
Figure~\ref{fig:fitDC} suggests a camera misalignment of approximately
$0.6^{\circ}$, which is reasonable for our experimental setup.
Even though the camera was misaligned, we were able to verify the flow
in the microchannel.

\subsection{Robustness Against Noisy Data}
In the past, when normalized velocities have been measured in a more
qualitative fashion~\cite{meinhart1999piv,sott2013mupiv}, interesting
features in velocity profiles, such as the kinks highlighted in
Figure~\ref{fig:revProc}, have typically been ignored.
In the more quantitative
evaluations~\cite{kloosterman2011flow,westerweel2004single,cierpka2012particle},
one has tried to eliminate any kinks or other features near walls
via post-processing meant to reduce the effects of a finite SV.

\begin{figure*}[htp]
\centering
\begin{subfigure}[t]{0.48\textwidth}
    \caption{\hfill~}
    \includegraphics[width=\textwidth]{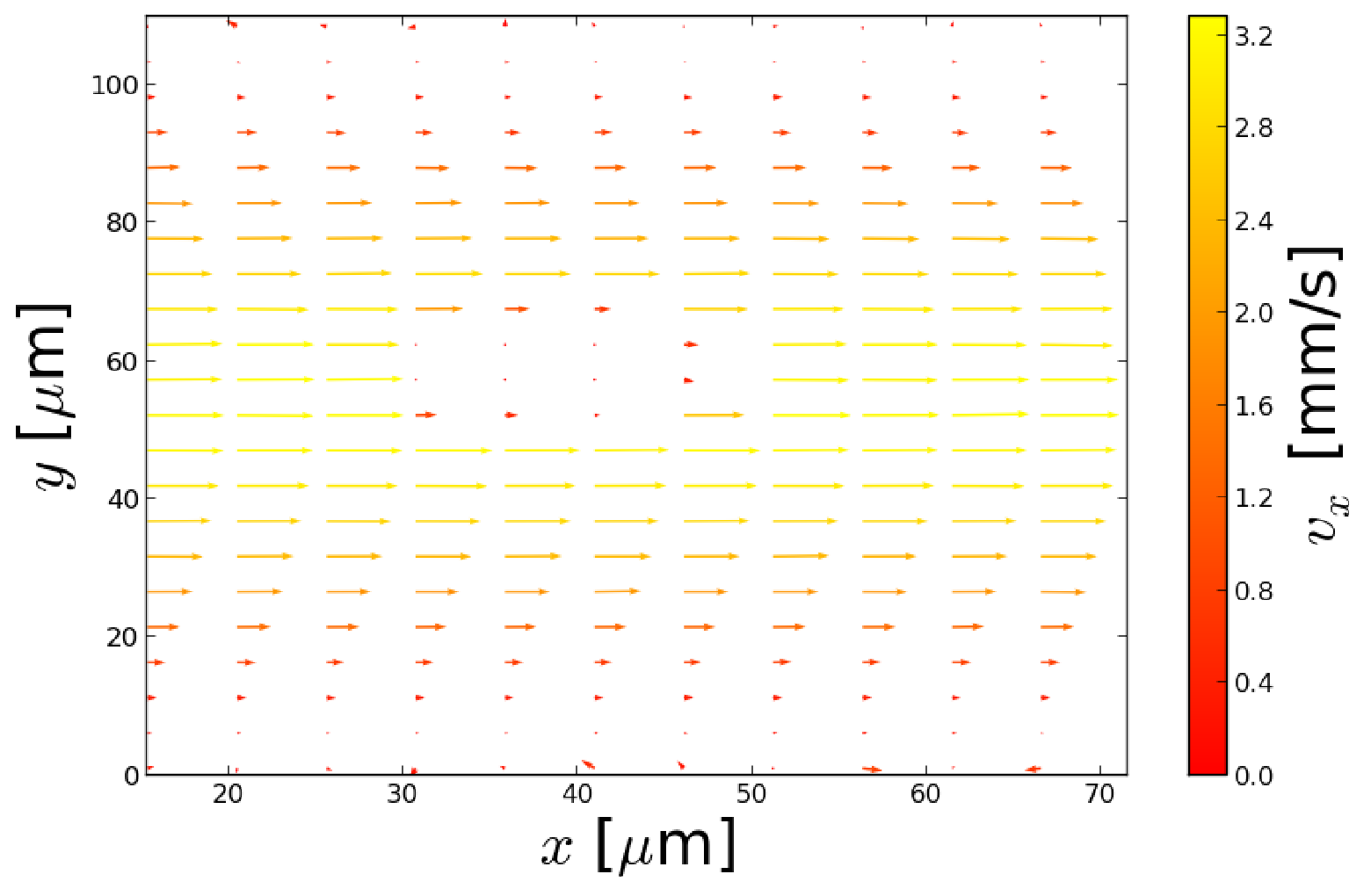}
    \label{fig:qhole}
\end{subfigure}
\hfill
\begin{subfigure}[t]{0.48\textwidth}
    \caption{\hfill~}
    \includegraphics[width=\textwidth]{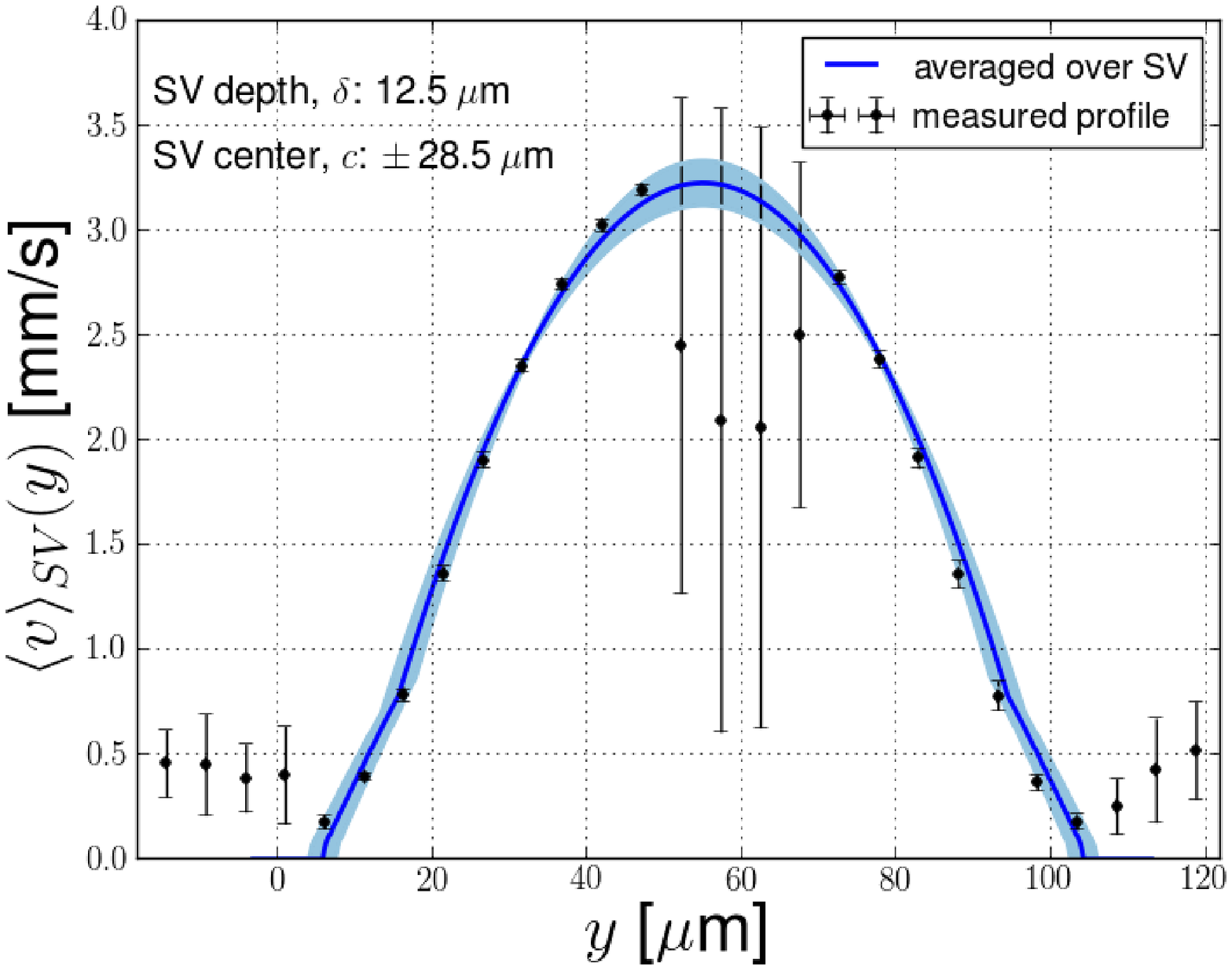}
    \label{fig:fhole}
\end{subfigure}
\caption{Flow profile for an experiment in which some of the velocity vectors
could not be determined correctly.
The pump rate for this experiment was 75\,$\mu$l/h. 
(\subref{fig:qhole}) Measured flow field in which at least four of the profiles
have incorrect velocities near the center of the channel.
(\subref{fig:fhole}) Fit to the data,
demonstrating the robustness of our method.
The large error bars near the center of the profile result
from the incorrect vectors.}
\label{fig:fithole}
\end{figure*}

With Figure~\ref{fig:fithole} we demonstrate how maintaining such
features can in fact lead to more robust $\mu$PIV measurements, particularly
when one has noisy data.
In this experiment, raw particle images included a large, stationary bright
spot that resulted, after processing, in the area with zero velocity shown in
Figure~\ref{fig:fithole}(\subref{fig:qhole}).
The average profile, shown in Figure~\ref{fig:fithole}(\subref{fig:fhole}), has
large uncertainty in the center of the channel.
Rather than discard the results of the experiment, by fitting the
entire measured speed profile, from wall to wall, we were able to quantify the
flow in the microchannel.
Because we fit the entire profile, and thereby incorporate more data
that have been minimally processed,
we believe that quantitative evaluation using the SV is significantly
more robust and more straightforward than with other methods.

\subsection{Varying Experimental and Processing Parameters}\label{sec:fitcompar}
In this section we show fits from a variety of other tests,
from which can be seen how variables such as flow rate, focus, and
processing options can affect the size, $\delta$, and position, $c$,
of the SV.
We also discuss how the measured SV differs from the
calculated DOC~\cite{olsen2000out}, and how the two have complimentary roles.
As in Section~\ref{sec:results}, error bars represent one standard deviation
in the set of all profiles measured along the $x$ direction.
The blue shaded area represents the combined uncertainty in the actual
fluid flow, which results from our uncertainty in the channel cross section
and the flow rate of the pump.
Note that, to within our experimental uncertainties, a simple
rectangular SV suffices to explain all of the results.

\begin{figure*}[htp]
\centering
\begin{subfigure}[t]{0.48\textwidth}
    \caption{\hfill~}
    \includegraphics[width=\textwidth]{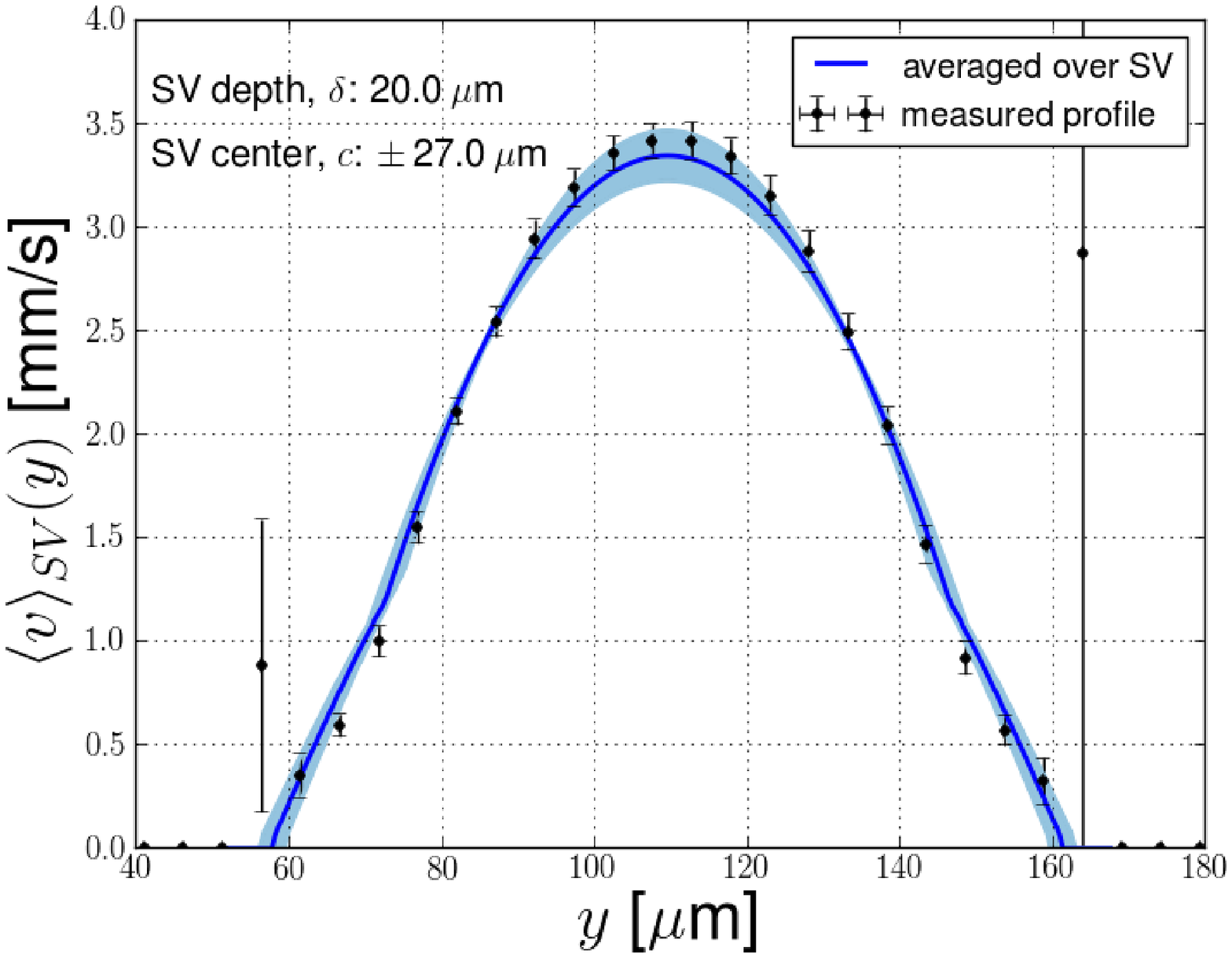}
    \label{fig:751}
\end{subfigure}
\hfill
\begin{subfigure}[t]{0.48\textwidth}
    \caption{\hfill~}
    \includegraphics[width=\textwidth]{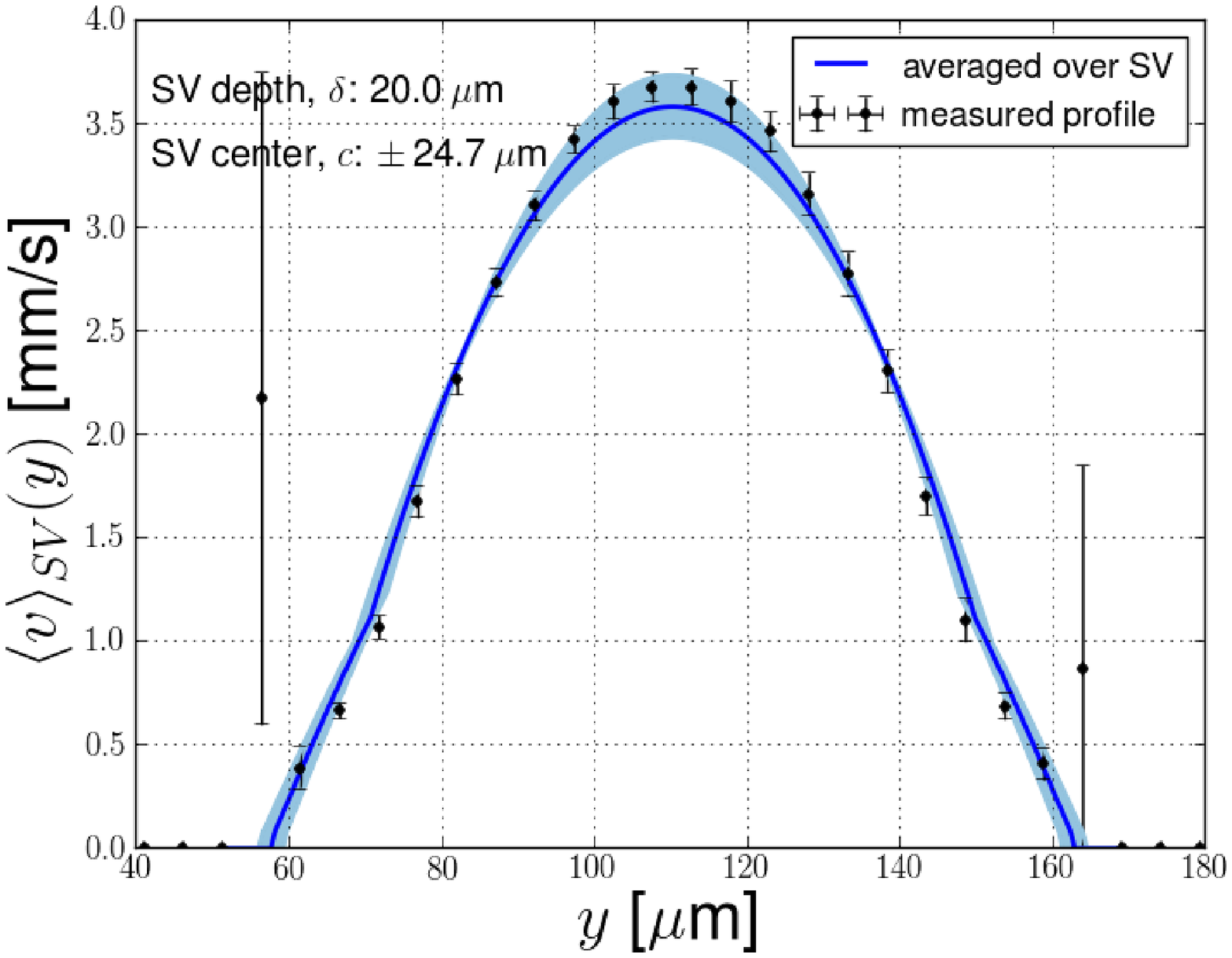}
    \label{fig:752}
\end{subfigure}
\caption{Fits to the experimental profiles obtained with fluid injection rate of
75\,$\mu$l/h.
(\subref{fig:751}) Before disconnecting from the pump and (\subref{fig:752})
after reconnecting to the pump and refocusing.}
\label{fig:c075}
\end{figure*}

Figure~\ref{fig:c075} shows fits for two measurements on the same
channel and for the same flow rate.
The velocity field of the fluid within the channel
is therefore the same for both measurements.
The velocity profiles measured by $\mu$PIV differ, however, because,
between the measurement in Figure~\ref{fig:c075}(\subref{fig:751})
and the measurement in Figure~\ref{fig:c075}(\subref{fig:752}), the
pump was briefly disconnected from and then reconnected to the microchannel,
and the microscope was refocused.
All processing options were consistent between the two experiments.
Though quality of focus is largely subjective, averaging over the SV
can explain both measurements.
In this case, refocusing changes the position of the SV but not its thickness.

\begin{figure*}[htp]
\centering
\begin{subfigure}[t]{0.48\textwidth}
    \caption{\hfill~}
    \includegraphics[width=\textwidth]{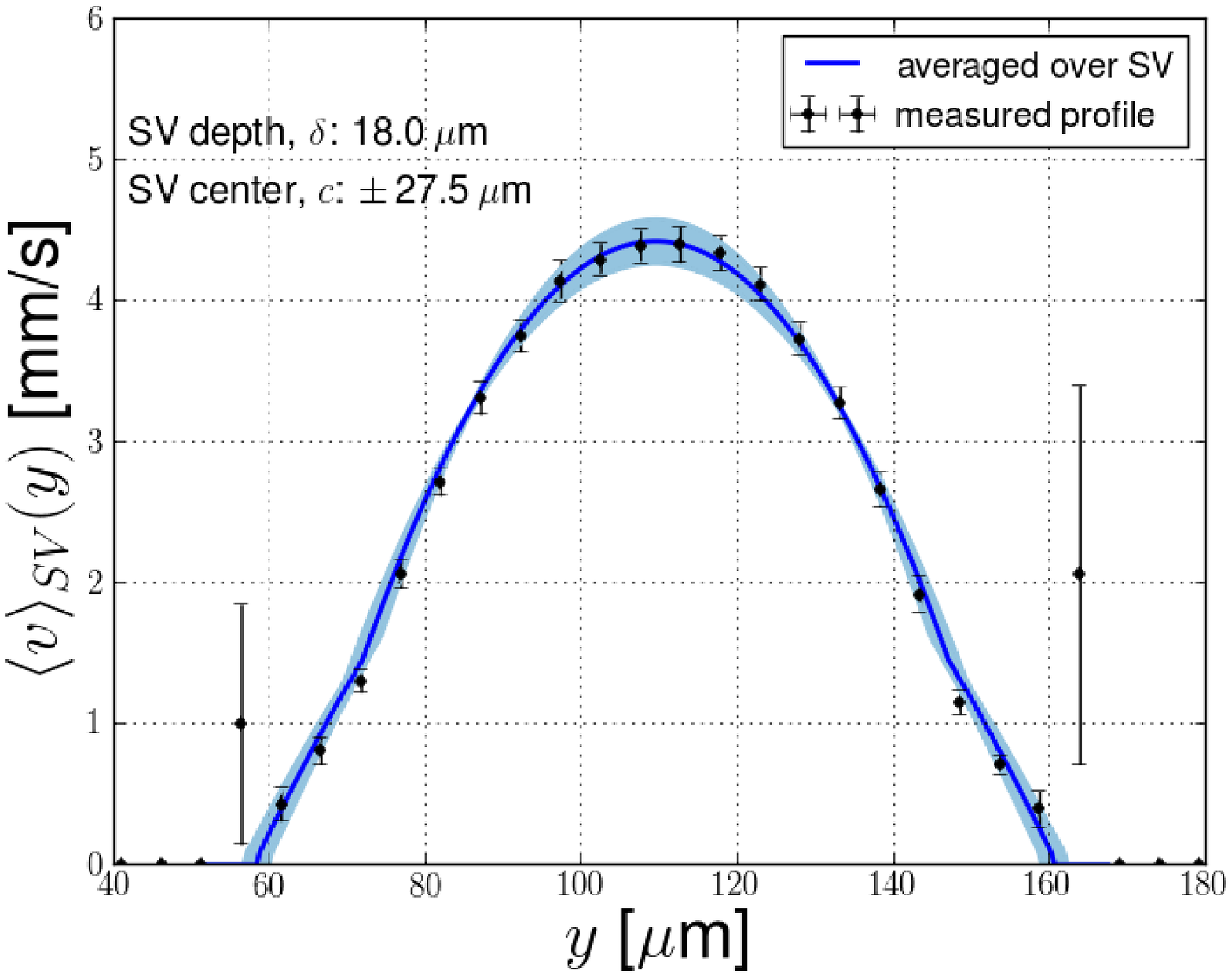}
    \label{fig:100norm}
\end{subfigure}
\hfill
\begin{subfigure}[t]{0.48\textwidth}
    \caption{\hfill~}
    \includegraphics[width=\textwidth]{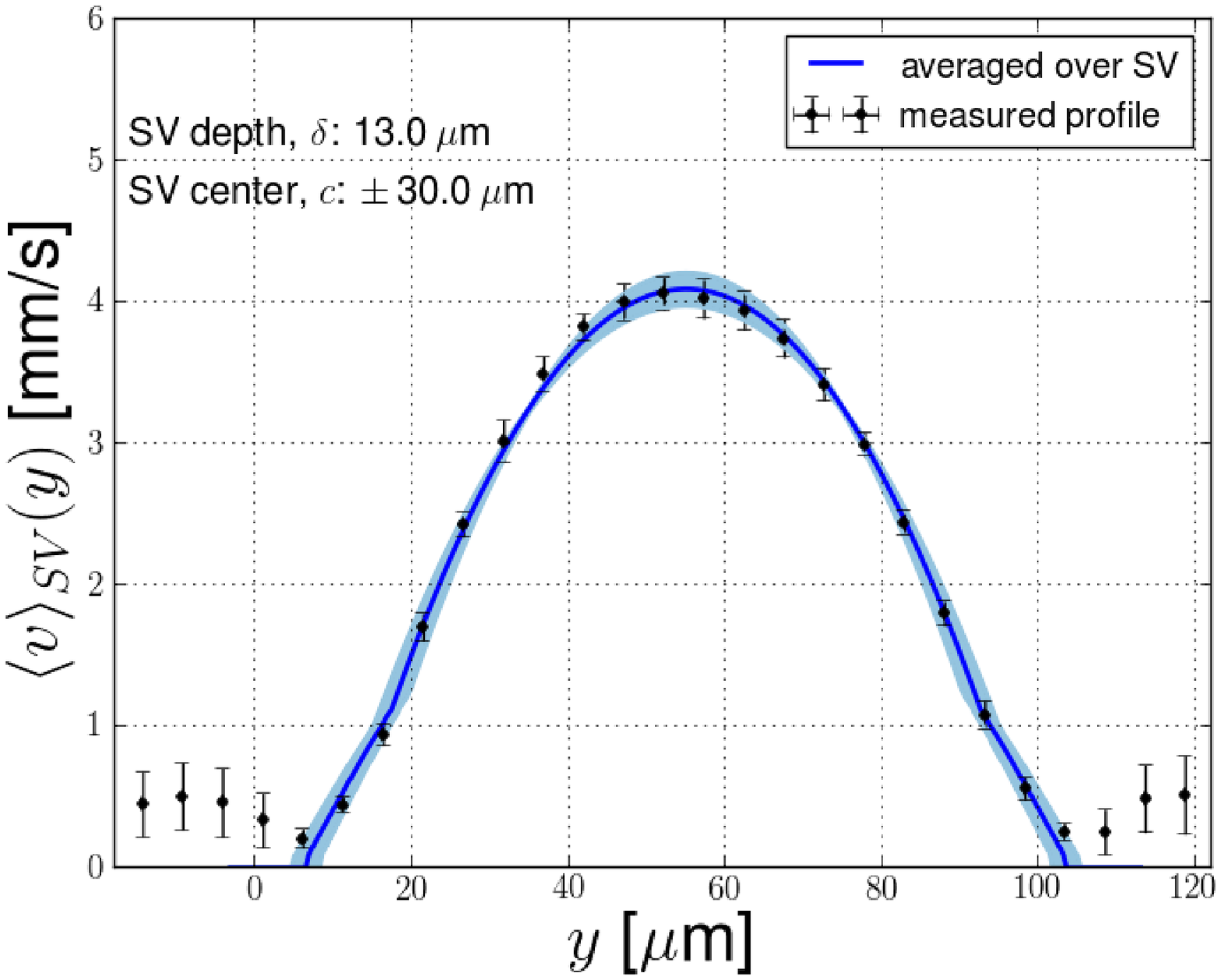}
    \label{fig:100Open}
\end{subfigure}
\caption{Flow profile obtained after processing the same set of CCD images with
(\subref{fig:100norm}) Insight 3G and (\subref{fig:100Open}) OpenPIV software.
The processing options are detailed in Section~\ref{sec:datproc}.
Figure~\ref{fig:open100}(\subref{fig:100norm})
is a reproduction of the fit shown in Figure~\ref{fig:cxFit}.
We have included it again here for easier comparison.}
\label{fig:open100}
\end{figure*}

In Figure~\ref{fig:open100} and Table~\ref{tab:comp} one can see the effects
on the SV of differing processing options and flow rates.
Figure~\ref{fig:open100} shows two measured results for the
same experiment, in which the fluid injection rate was 100\,$\mu$l/h.
Both measured velocity profiles were obtained from the same set of CCD images.
They differ only in the image processing options, which are discussed in
Section~\ref{sec:datproc}.
The positions of the SV center, $c$, and the SV thickness, $\delta$,
resulting from the various processing options are listed in
Table~\ref{tab:comp} for two fluid injection rates:
100\,$\mu$l/h and 75\,$\mu$l/h.
For both flow rates, processing with OpenPIV results in a substantially 
thinner SV that is located slightly farther from the center of the channel.
The preferred processing options may vary from case to case, and it is
important to note that their effects on the SV can be quantified.
\begin{table}[h]
\centering
\begin{tabular}{l|l|l|}
\cline{2-3}
                                            & \multicolumn{1}{c|}{Insight 3G} & \multicolumn{1}{c|}{OpenPIV} \\ \hline
\multicolumn{1}{|l|}{\multirow{2}{*}{100\,$\mu$l/h}} & $\delta=18.0\,\mu$m  & $\delta=13.0\,\mu$m   \\ \cline{2-3} 
\multicolumn{1}{|l|}{}                      & $c=\pm 27.5\,\mu$m  & $c=\pm 30.0\,\mu$m   \\ \hline
\multicolumn{1}{|l|}{\multirow{2}{*}{75\,$\mu$l/h}} & $\delta=20.0\,\mu$m  & $\delta=12.5\,\mu$m   \\ \cline{2-3} 
\multicolumn{1}{|l|}{}                      & $c=\pm 27.0\,\mu$m  & $c=\pm 29.0\,\mu$m   \\ \hline
\end{tabular}
\caption{Comparison of the SV for two processing options and flow rates.}
\label{tab:comp}
\end{table}

As discussed in Section~\ref{sec:intrsv}, the DOC derived
in~\cite{olsen2000out} is often used as an estimate for the thickness of the
SV, $\delta$.
We calculate the DOC for our experiments to be approximately $37\,\mu$m.
This value is based on an assumed intensity threshold, $\varepsilon$,
required for a particle image to affect significantly the processing.
We have used the typical~\cite{cierpka2012particle} value
of $\varepsilon=0.01$, which means we assume that contributions are made by
out-of-focus particles with a minimum of 1\% of the intensity of an
in-focus particle.
Note that this calculated value of the DOC is constant for all of our
experiments presented in Sections~\ref{sec:results} and~\ref{sec:applications},
for which we have measured that the actual $\delta$ varies
between $12.5\,\mu$m and $20\,\mu$m.
The calculated DOC can be reconciled with the measured $\delta$ if one
assumes that $\varepsilon$ in fact varies between $\varepsilon=0.245$
and $\varepsilon=0.075$, but it is not clear how one could determine
these thresholds without first measuring the SV for the various focus levels
and processing options.
While the DOC gives important intuition into the variables affecting
$\delta$, it does not replace a quantitatively determined SV.

\subsection{Analysis of Scanning PIV}\label{sec:prevwork}
Kloosterman et al.~\cite{kloosterman2011flow} have recently claimed that
a simple average, such as Equation~\eqref{eq:vavedoc}, cannot
explain the velocity measured by $\mu$PIV with a large depth of field.
Rather than treat the entire profile from wall to wall,
they have considered the difference between measured and predicted
maximum speeds at the center of the microfluidic channel.
In our Figure~\ref{fig:cxFit}, for example, this corresponds to explaining
why the maximum measured speed is
approximately 4.4\,mm/s (in symbols) rather than the theoretical maximum
for a centered SV with zero thickness, 6.4\,mm/s (dashed line).

Their claim is based on a common experimental procedure called ``Scanning PIV'',
in which one systematically varies the location of the focal plane
along the optical ($z$) axis and measures the
velocity profile for each location.
Their analysis is for Hagen-Poiseuille flow in a cylindrical channel,
and because
the maximum velocity in the channel occurs at its vertical center, the
authors claim that from all the measured profiles from the scan,
the one with the highest peak velocity is the centered profile.
Having centered the region over which particle images are resolved, then,
they find that for large depths of field that cover the entire
channel, the measured velocity is larger than what is predicted from
averaging over the entire channel.

Using the average over the SV in Equation~\eqref{eq:vavedoc}, however, we can
show that this use of Scanning PIV will not find the center of the 
channel for large SV.
To analyze the results published in Ref.~\cite{kloosterman2011flow},
we use the Hagen-Poiseuille profile for flow in a cylindrical channel,
\begin{equation} \label{eq:poise}
v_P(r) = v_{0}\left(1-\frac{r^2}{R^2}\right),
\end{equation}
where $v_{0}$ is the maximum speed in the channel,
$r$ is the radial coordinate, $R$ is the channel radius, and
$r=0$ at the center of the channel.
Let us change to Cartesian coordinates with $z=0$ at the vertical center
of the channel, and substitute $D=2R$ for the channel's diameter.
In this section we are interested in the maximum speed,
which occurs at the fixed $y$ value of the center of the channel,
$y=y_{\textrm{center}}$, while the SV changes.
We will therefore fix $y$ and emphasize the dependence of the speed on the
position, $c$, and size, $\delta$, of the SV.
Let $\langle v\rangle_{SV}(y=y_{\textrm{center}})\equiv
\langle v\rangle_{SV}(c,\delta)$ denote our prediction of the maximum speed.
Using Equation~\eqref{eq:vavedoc} with the Poiseuille profile~\eqref{eq:poise},
we find
\begin{equation} \label{eq:vimproved}
\langle v\rangle_{SV}(c,\delta) = v_{0} \left[ 1-\frac{4}{3} \left(
\frac{z_{\mbox{\tiny --}}^2+z_{\mbox{\tiny --}}z_{\mbox{\tiny +}}
+z_{\mbox{\tiny +}}^2}{D^2} \right) \right],
\end{equation}
where the dependence on $c$ and $\delta$ is in the integration
bounds
$z_{\mbox{\tiny --}} = \textnormal{Max} \left(c-\delta/2,-D/2\right)$ and
$z_{\mbox{\tiny +}} = \textnormal{Min} \left(c+\delta/2,D/2\right)$, as before.

In terms of the notation in Equation~\eqref{eq:vimproved}, when one uses
scanning PIV to find the center of a channel for a given $\delta$,
one varies $c$ until $\langle v\rangle_{SV}(c,\delta)$ is maximized.
The principal assumption, then, is that this maximum always occurs at $c=0$.
To check this assumption, we solve
$\frac{\partial}{\partial c} \langle v\rangle_{SV}(c,\delta) = 0$ for $c$
and find multiple solutions.
One solution is for a centered SV, with $c=0$, for which the maximum speed is
\begin{equation} \label{eq:centeredDOC}
\langle v\rangle_{SV}(c,\delta) \Big|_{c=0} =
\begin{cases}
v_{0}\left(1-\frac{\delta^2}{3 D^2}\right) & 0\leq\delta < D \\
\frac{2}{3}v_{0} & \delta \geq D.
\end{cases}
\end{equation}
Another is for $c=\pm(D-2\delta)/4$, for which the SV is not centered in the
channel, and the maximum speed is
\begin{equation} \label{eq:offcenteredDOC}
\langle v\rangle_{SV}(c,\delta) \Big|_{c=\pm(D-2\delta)/4} =
\begin{cases}
v_{0}\frac{(9D^2+12D\delta-16\delta^2)}{12D^2}
& 0\leq\delta < \frac{3}{4}D \\
\frac{3}{4}v_{0} & \delta \geq \frac{3}{4}D.
\end{cases}
\end{equation}

\begin{figure}[htp]
\begin{center}
\includegraphics[width=.5\textwidth]{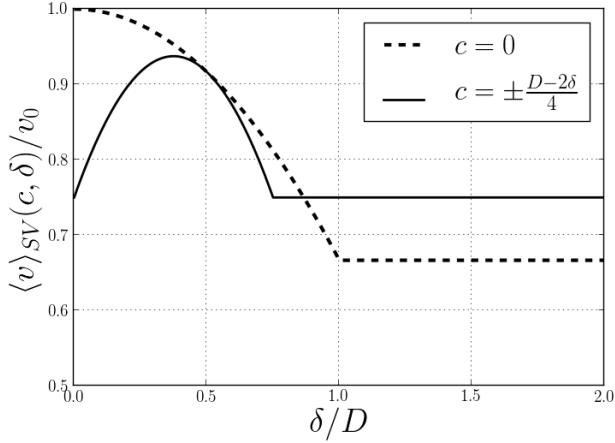}
\end{center}
\caption{Maximum speed measured by the scanning PIV procedure as a function of
$\delta$.
The dashed line is for a DOC centered at $c=0$, while the solid line
represents the maximum speed when the DOC is centered at $c=\pm(D-2\delta)/4$.}
\label{fig:scanpvim}
\end{figure}

Figure~\ref{fig:scanpvim} is a plot of $\langle v\rangle_{SV}(c,\delta)$ for
these two maximizing values of $c$.
The curves cross at $\delta=\frac{\sqrt{3}}{2}D$.
Because the experimenter selects the value of $c$ that maximizes
the measured speed, for $\delta>\frac{\sqrt{3}}{2}D$ in a cylindrical channel,
this scanning PIV method will position the SV away from the channel's center,
at $c=\pm(D-2\delta)/4$.
Thus, one must use Scanning PIV with caution.

Finally, we note that in Ref.~\cite{kloosterman2011flow}
the authors base their claim on a method for calculating
the simple average that is different from Equation~\eqref{eq:vavedoc}.
Because of the assumed ability of scanning PIV to locate the center of
the cylindrical channel, they fix $c=0$.
Furthermore, they do not stop the integration at the bottom and top of
the channel, and instead use
$z_{\mbox{\tiny --}} = -DOC/2$ and $z_{\mbox{\tiny +}} = +DOC/2$, where
$DOC$ is the Depth of Correlation we have discussed in
Section~\ref{sec:fitcompar}.
The result is
\begin{equation} \label{eq:kloos1}
v_{\textnormal{max}}(DOC)=v_{0}\left(1-\frac{DOC^2}{3D^2}\right),
\end{equation}
which has positive values only for $DOC/D<\sqrt{3}$.
Though the $DOC$ does not equal $\delta$, the two can be roughly compared
when both are as large as or larger than the channel diameter, $D$, and
images cover the entire depth of the channel.
In this case, the authors predict from Equation~\eqref{eq:kloos1} that
$v_{\textnormal{max}}(D)=\frac{2}{3} v_{0}$,
but they measure $v_{\textnormal{max}}(D)\approx \frac{3}{4} v_{0}$ and 
find that it remains nearly constant for $DOC \gtrsim D$.
We see from Figure~\ref{fig:scanpvim} and Equation~\eqref{eq:offcenteredDOC},
however, that the simple average, combined with this use of Scanning PIV,
does predict
$\langle v\rangle_{SV}(c,\delta)=\frac{3}{4}v_{0}$ for large $\delta$.

\section{Conclusion}
We show that a simple average over the Sampling Volume (SV) suffices to explain 
quantitatively the velocity profiles measured in $\mu$PIV experiments for
which the depth of field has a size comparable to that of the
microfluidics system itself.
This allows rigorous verification of the velocity fields
of fluids flowing in microchannels.
With this simple approach, one can easily understand and
physically interpret the measured features of velocity profiles.
In some cases, it allows the complete determination of an unknown flow rate
or velocity field based only on the $\mu$PIV measurement.
Furthermore, in a straightforward manner one can determine measurement
uncertainty, which is essential for any quantitatively useful experiment.
Because of the presence of a finite sampling region,
in the past one has either ignored the magnitude of the measured velocity
or else treated the effects of the finite volume as errors to be
corrected by complicated processing.

By minimally processing data and retaining all measured features,
such as kinks in the velocity profile, we have developed a method that
is especially robust against noise and other common sources of
measurement error.
By comparing the measured velocity profiles to the theoretical predictions,
we are able to quantify the effect on the SV of processing options and
focus.
We also provide a critical evaluation of Scanning PIV, which is often used to
locate the center of a microchannel.
We show how and why it in fact fails to locate the center for large SV.
In general, the concept of the SV has implications for quantitative
measurement of flow within channels on the micrometer scale and below. 

\section*{Acknowledgments}
We would like to thank Michael Engel from IBM Research -- Watson as well as the
LABNANO--CBPF for the SEM images,
Diney Ether and the LPO--UFRJ for help with the calibration,
Angelo Gobbi at the LMF--LNNano for profilometer measurements, and
Jos\'{e} Flori\'{a}n at PUC--Rio for help with the $\mu$PIV equipment.

\singlespacing
\begin{footnotesize}
\bibliographystyle{unsrt}
\bibliography{biblioPIV}
\end{footnotesize}
\end{document}